\documentclass[prl,aps,english,showpacs,twocolumn,superscriptaddress]{revtex4-1}
\usepackage[english]{babel}
\usepackage{amsfonts}
\usepackage{verbatim}
\usepackage{amsmath}
\usepackage{amsthm}
\usepackage{amssymb}
\usepackage{graphicx}
\usepackage{subfigure}
\usepackage{cancel}
\usepackage[usenames,dvipsnames]{color}
\newcommand{\um} {~\mu\mathrm{m}}

\newcommand{\ludo}[1]{{\color{black} #1}}
\renewcommand{\phi} {\varphi}
\usepackage{natbib}

\definecolor{myc}{RGB}{0,0,0}

\begin{document}

\title{The glass transition of soft colloids}

\author{Adrian-Marie Philippe}
\email{adrian-marie.philippe@umontpellier.fr}

\author{Domenico Truzzolillo}
\affiliation{L2C Univ Montpellier, CNRS, Montpellier, France}

\author{Julian Galvan-Myoshi}
\affiliation{University of Fribourg, Department of Physics, CH-1700 Fribourg}

\author{Philippe Dieudonn\'e-George}
\affiliation{L2C Univ Montpellier, CNRS, Montpellier, France}

\author{V\'eronique Trappe}
\affiliation{University of Fribourg, Department of Physics, CH-1700 Fribourg}

\author{Ludovic Berthier}
\affiliation{L2C Univ Montpellier, CNRS, Montpellier, France}

\author{Luca Cipelletti}
\affiliation{L2C Univ Montpellier, CNRS, Montpellier, France}


\date{\today}

\begin{abstract}
We explore the glassy dynamics of soft colloids
using microgels and charged particles interacting by
steric and screened Coulomb interactions, respectively. In the supercooled regime, the structural relaxation time $\tau_\alpha$ of both systems grows steeply with volume fraction, reminiscent of the behavior of colloidal hard spheres. Computer simulations confirm that the growth of $\tau_\alpha$ on approaching the
glass transition is independent of particle softness. By contrast, softness becomes relevant at very large packing fractions when the system
falls out of equilibrium. In this non-equilibrium regime, $\tau_\alpha$  depends surprisingly weakly on packing fraction and time
correlation functions exhibit a compressed exponential decay consistent with
stress-driven relaxation. The transition to this novel regime
coincides with the onset of an anomalous decrease of local order
with increasing density \ludo{typical of ultrasoft systems. We propose that these peculiar dynamics results from the combination of the  non-equilibrium aging dynamics expected in the glassy state and the tendency of colloids interacting through soft potentials
to refluidize at high packing fractions.}
\end{abstract}

\maketitle

%
%

The dramatic slowing down of the structural relaxation time upon modest variations of a control parameter is a general phenomenon observed in a wide range of glass-formers, ranging from molecular systems~\cite{debenedetti2001,BB11}, to soft matter~\cite{weeks_review} and biological systems~\cite{angelini}.
For colloidal systems, the relevant control parameters are the particle volume fraction $\phi$ and the strength of the
interparticle interactions. Because hard sphere interactions are central to
theoretical and computational studies that capture the essence of the glass transition~\cite{BB11}, hard sphere colloidal systems have been
extensively investigated~\cite{weeks_review}.

The glass transition of hard sphere systems is well documented.
With increasing $\phi$,
positional correlations develop, reflected by
the appearance of a diffraction peak in the static structure factor $S(q)$
at a wavevector $q_{\rm max}$ corresponding to the
typical interparticle distance.
Both $q_{\rm max}$ and $S(q_{\rm max})$ monotonically increase with $\phi$,
whereas the structural relaxation time $\tau_{\alpha}$ measured, e.g.,
in scattering experiments increases by several orders of
magnitude. The \ludo{initial growth can be} described as a power law divergence, $\tau_{\alpha} \propto (\phi_{\rm mct}-\phi)^{-\gamma}$~\cite{pusey,vanMegen}, consistent with mode-coupling theory~\cite{BGS84}. \ludo{The data at deep supercooling are better fitted  to an exponential divergence, $\tau_\alpha \sim \exp[A/(\phi_0-\phi)^\delta]$, where typically $\phi_0 > \phi_{\rm mct}$~\cite{BEPPSBC08}.}
This exponential growth bears some (formal~\cite{BW09})
resemblance with the fragile behaviour of molecular glass-formers~\cite{mattsson2009}.
Compressing hard spheres further, the system enters a
non-equilibrium aging regime, where the relaxation time
increases \ludo{rapidly} with the age of the material~\cite{masri2009},
until it becomes so large that no relaxation is measurable on accessible time scales.

The focus recently shifted from hard \ludo{colloids to a large variety of soft colloidal particles, such as emulsions, microgel suspensions or biological systems,} in view of their interest for both fundamental science and applications~\cite{yunker}. Soft colloids can overlap and deform and may thus
be compressed up to packing fractions that cannot be explored with hard
particles. Two striking signatures of particle softness were
reported. First, softness results in an anomalous non-monotonic
evolution of $S(q_{\rm max})$ with $\phi$,
which initially increases as in hard spheres, but then
decreases at larger
$\phi$~\cite{ZhangNat2009,graves2005,caswell2013,paloli2013,seekell2015},
as a result
of a competition between entropy and energy~\cite{jacquin}. Theory
suggests that this loss of local order at large
$\phi$ is accompanied by a reentrant glass transition
and a complete suppression of aging~\cite{berthier2010pre},
\ludo{reported in Ref.~\cite{srivastava2013}}. Second, it was argued~\cite{senff_temperature_1999,mattsson2009,seekell2015,gupta_validity_2015,Nigro_Angelini_Bertoldo_Bruni_Ricci_Ruzicka_2017}
that softness changes the nature of the glassy
dynamics. In particular, a very gradual increase
of the relaxation time of the form $\tau_{\alpha} \sim \exp(B\phi)$ was
reported~\cite{mattsson2009,Nigro_Angelini_Bertoldo_Bruni_Ricci_Ruzicka_2017,yang},
in stark contrast with hard
sphere behaviour and other studies of soft particles~\cite{BW09,Xu09,Vagberg11,pellet_glass_2016,li_long-term_2017}. These conflicting reports show the lack of consensus about how
softness impacts the dynamical slowing down with \ludo{density,}
and how structural and dynamical anomalies relate to each other.

In this work, we provide a coherent picture of the behaviour of \textcolor{myc} {colloidal particles interacting via a soft, repulsive potential, by determining experimentally
the $\phi$-dependent structural, dynamical and rheological properties
of soft colloids. \ludo{We support our results using a simple numerical model of soft particles, where the magnitude of soft repulsion can be easily tuned over a very broad range and its impact on the equilibrium glassy dynamics analysed carefully.}
Previous experiments predominantly focused on microgel particles~\cite{senff_temperature_1999,mattsson2009,ZhangNat2009,Nigro_Angelini_Bertoldo_Bruni_Ricci_Ruzicka_2017,li_long-term_2017,van_der_scheer_fragility_2017} formed by permanently crosslinked polymer chains. While microgels are a convenient experimental realization of soft particles~\cite{yunker}, their polymeric nature introduces additional degrees of freedom and complexity, making data interpretation difficult: whether microgels deform~\cite{seth_micromechanical_2011,mattsson2009} or interpenetrate~\cite{mohanty_interpenetration_2017} at high volume fraction is still a debated issue, as is the role of entanglements and chain relaxation in the observed dynamics~\cite{li_long-term_2017}. In our work, we overcome these difficulties by systematically comparing the behavior of microgels to that of compact silica particles interacting via a soft potential, for which no polymeric degrees of freedom are present. \ludo{This allows us to disentangle unambiguously the role of particle softness from any other effect.}} The microgels are poly(N-isopropylacrylamide)
(PNiPAM) microgels synthesized as in~\cite{SenffJCP1999}, which we produce in two batches with hydrodynamic diameter of respectively $d_h = 353$ nm and $d_h = 268$ nm at temperature $T = 293$~K. The silica particles are Ludox-TM 50 (Sigma-Aldrich), with $d_h = 46$ nm, see~\cite{SupInfo} for details on the samples and their preparation.
Silica particles are \textcolor{myc} {compact,} hard and undeformable, but they interact through a soft repulsive Yukawa potential, due to their surface charge.
The static structure factor of the suspensions is obtained either by static light scattering (SLS, for the microgels) or by small-angle X ray scattering (SAXS, for the silica particles), where $q = 4\pi n \lambda^{-1} \sin(\theta/2)$ is the scattering vector, with $\lambda = 532.5$ (resp., $0.154~\mathrm{nm})~$ the wavelength of the incident laser (resp., X-ray) radiation, $n$ the solvent refractive index and $\theta$ the scattering angle. The macroscopic flow properties are measured by rheology, using a stress-controlled rheometer~\cite{SupInfo}. The microscopic dynamics are probed by dynamic light scattering (DLS~\cite{Berne1976}), using a commercial apparatus for diluted samples and custom setups~\cite{el_masri_aging_2005,philippe2016} with a CMOS detector for concentrated suspensions. Most of the DLS data are collected at $\theta = 90^{\circ}$ ($q = 22.19\um^{-1}$), but we also perform experiments at $\theta = 180^{\circ}$ ($q = 31.39\um^{-1}$). DLS experiments yield the two-time intensity autocorrelation function $g_2(q;t_w,\tau)-1$ describing the relaxation of density fluctuations of wavevector $q$, as a function of sample age $t_w$ and delay time $\tau$~\cite{SupInfo}. The intensity correlation function is related to the intermediate scattering function $f(q; t_w, \tau)$ by $g_2-1 = f^2$. We use molecular dynamics to simulate soft repulsive
particles interacting via a harmonic potential, as detailed in
Ref.~\cite{BW09}.
The harmonic potential is a good model for soft microgels~\cite{harmonic_yodh}. \ludo{The model neglects polymeric degrees of freedom. Its physics is controlled by the particle softness, which can be tuned at will, and which is expressed by the ratio
$\tilde{\epsilon} = \epsilon / (k_B T)$ between
the elastic energy scale, $\epsilon$, and thermal fluctuations, $k_B T$.} The system behaves as
nearly hard spheres when $\tilde{\epsilon}
> 10^6$, whereas soft microgels typically have
$\tilde{\epsilon} \approx 10^3$~\cite{harmonic_yodh,ikeda_disentangling_2013}.
\ludo{Simulations are used to understand the impact of particle softness on the equilibrium glassy dynamics, and we do not explore the aging regime numerically.}

\begin{figure}
\includegraphics[width=1.\columnwidth]{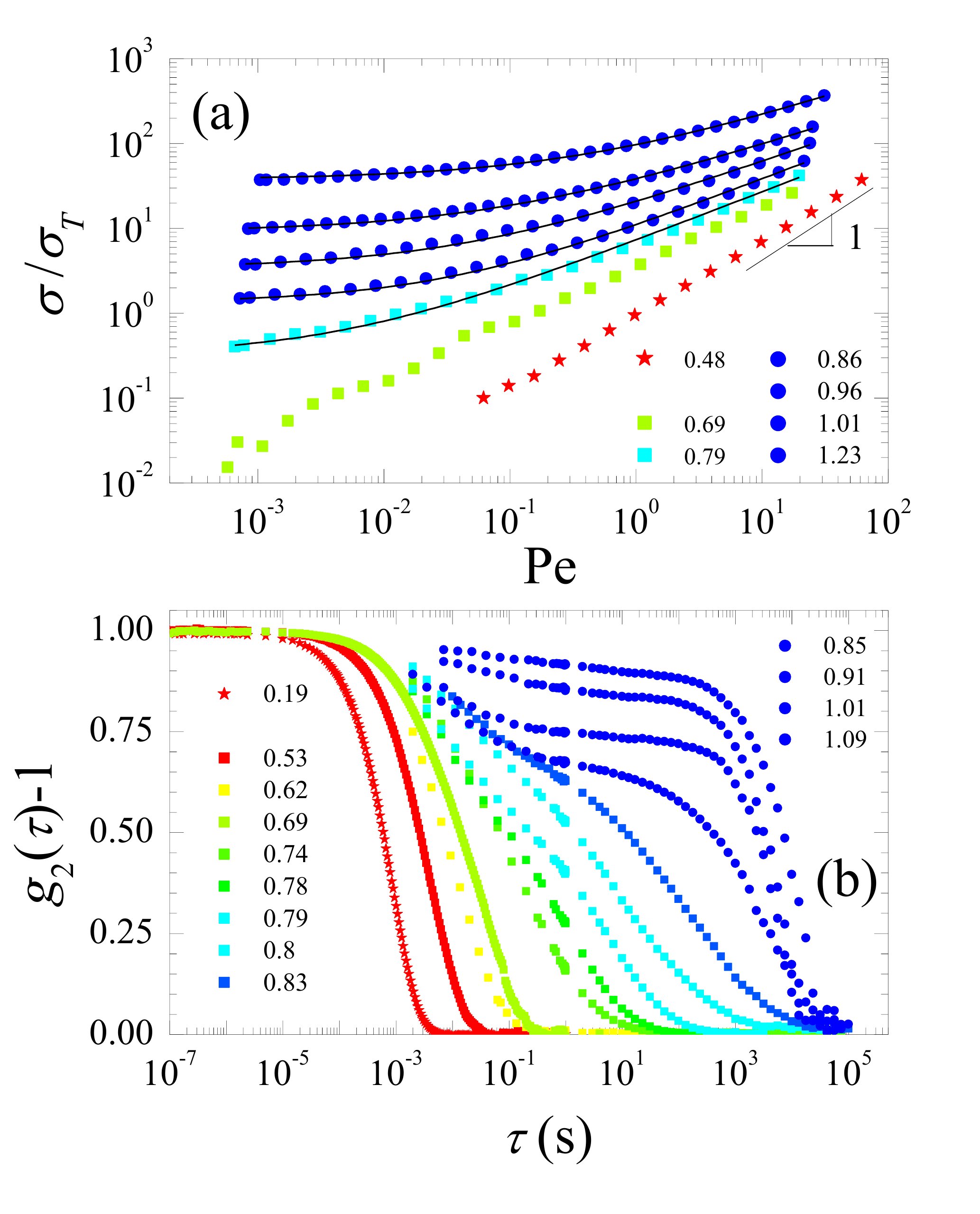}
 \caption{(a) Representative flow curves of PNiPAM suspensions
with stars, squares and circles respectively corresponding to the
regimes I, II and III described in the text. Lines are Herschel-Bulkley fits.
(b) Representative correlation functions of PNiPAM suspensions, with
symbols chosen as in (a).}
 \label{FlowPNiPAM}
\end{figure}

Selected examples of the $\phi$-dependent flow curves are shown in
Fig.~\ref{FlowPNiPAM}(a), where shear stress $\sigma$ is normalized by the entropic stress of dense Brownian suspensions,  $\sigma_T = k_BT/d_h^3$, and the shear rate $\dot{\gamma}$ is normalized by the Brownian diffusion rate, which results in the P\'eclet number, $Pe = 3\pi\eta_s\dot{\gamma}d_h^3/(k_BT)$; $k_B$ is the Boltzmann constant.
Newtonian behaviour, $\sigma \propto \dot{\gamma}$, is observed at low $\phi$. With increasing $\phi$ we start to observe shear-thinning, and for
$\phi \ge 0.79$ we find that the
flow curves are well-fitted by the Herschel-Bulkley
law~\cite{Bonn_Denn_Berthier_Divoux_Manneville_2017}, $\sigma=\sigma_y+a Pe^{b}$. This
signals the emergence of a finite yield stress $\sigma_y$
and thus marks a transition from a fluid to an amorphous
solid~\cite{Bonn_Denn_Berthier_Divoux_Manneville_2017,petekidis2004}. Near the transition,
we find $\sigma_y / \sigma_T \approx {\cal O}(1)$, as observed
for Brownian hard spheres~\cite{petekidis2004}.
This is a first indication that the glass transition of microgels
is driven by entropy, as for hard spheres, rather than
by particle elasticity~\cite{ikeda_unified_2012,ikeda_disentangling_2013}.
The flow curves for silica particles exhibit a similar behavior~\cite{SupInfo}, except that a measurable yield stress emerges
at much lower volume fraction, $\phi \approx 0.395$,
confirming the key role played by long-range repulsion.

\begin{figure}
 \includegraphics[width=1.\columnwidth]{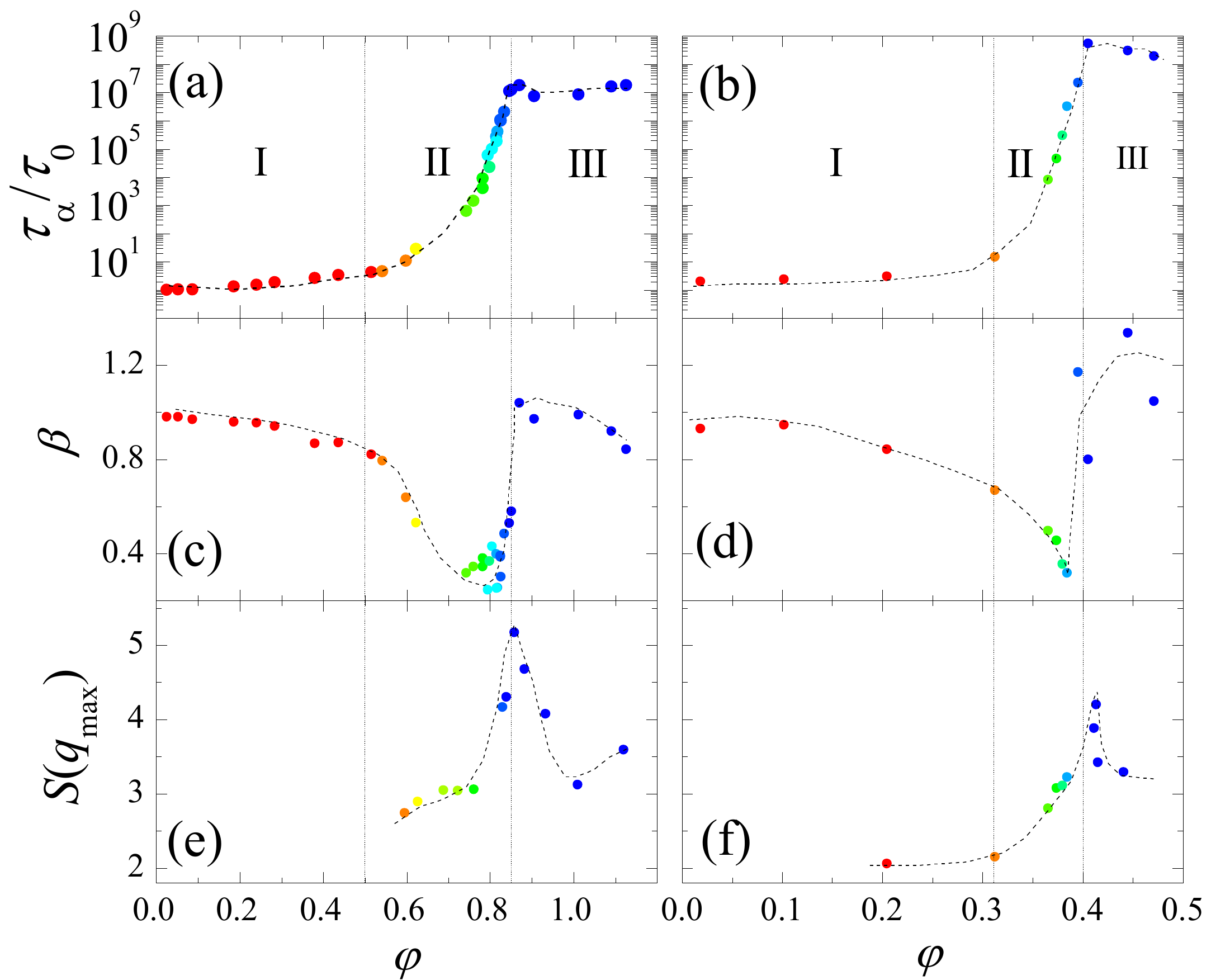}
 \caption{Volume fraction dependence of the relaxation
time (scaled by its dilute limit $\tau_0$) (a, b),
the stretching exponent of the scattering function (c, d),
and the height of the first peak of the structure factor (e, f).
Left column are data for PNiPAM, right column for Ludox.
The dashed lines in a)-f) are guides to the eye, the vertical lines indicate the approximate boundaries between the
different dynamical regimes.}
 \label{fig:two_systems}
\end{figure}

The evolution of the microscopic dynamics across the transition to solid-like behavior is shown
in Fig.~\ref{FlowPNiPAM}(b), where we display the data obtained for the PNiPAM samples. The correlation functions exhibit distinct characteristics,
which suggest the existence of three regimes.
In regime I, corresponding to $\varphi<0.5$, the dynamics are fast
and $g_2-1$ decays exponentially, with a decay time close to $\tau_0$, the relaxation time in the dilute limit.
In regime II, $0.5 \leq\varphi < 0.85$, the dynamics
slow down dramatically with increasing $\phi$. The relaxation time obtained from a
stretched exponential fit to the final decay of the correlation function,
$g_2 -1 \propto \exp [ - 2\left(\tau/\tau_\alpha\right)^\beta]$,
increases by 7 orders of magnitude (Fig.~\ref{fig:two_systems}(a)). Concurrently, the shape of $g_2-1$
becomes stretched, $\beta<1$, as shown in Fig.~\ref{fig:two_systems}(c).
The emergence of a yield stress
near $\phi\approx 0.8$ is accompanied by the onset of
caging, as signaled by the intermediate-time plateau in
$g_2-1$. Regime III is entered at higher packing fractions,
where the decay of the correlation functions becomes
much steeper, as shown by the rapid growth of $\beta$ up to values $\ge 1$, see Figs.~\ref{fig:two_systems}(c, d).
While the plateau height keeps increasing with $\phi$ (Fig.~\ref{FlowPNiPAM}(b)), indicating a
tighter particle caging,
the final relaxation time is weakly sensitive to $\phi$,
in stark contrast with regime II, see Fig.~\ref{fig:two_systems}(a).
\textcolor{myc} {\ludo{A similar saturation} of the relaxation time at very high $\phi$ has been reported very recently in PNiPAM-grafted polystyrene particles~\cite{li_long-term_2017}, \ludo{and} was attributed to the relaxation of the polymer shell. Crucially, we find that the same behaviour is observed for the Ludox suspensions~\cite{SupInfo},
see Figs.~\ref{fig:two_systems}(b, d). This rules out the polymeric nature of PNiPAM as an \ludo{explanation} and suggests that the scenario emerging from Fig.~\ref{FlowPNiPAM} is instead a more general feature of \ludo{soft colloidal particles.}}
Remarkably, our data in regime II do not display the gradual
(or `strong') increase of $\tau_\alpha$ reported in Ref.~\cite{mattsson2009}
for microgels, but a very steep (or `fragile') increase.

The transition between regimes II and III is characterized by the saturation of $\tau_\alpha$ and by a marked
minimum of the stretching exponent $\beta$, which first decreases to $\beta \approx 0.4$ but then steeply increases to $\beta \ge 1$, indicative of compressed exponential relaxation. Remarkably, the sharp dynamic crossover between regimes II and III is reflected in the static  structure factor.
The magnitude of the peak of the structure factor, $S(q_{\rm max})$, evolves non-monotonically
with $\phi$; it exhibits a maximum close to the
transition between the two regimes, see Figs.~\ref{fig:two_systems}(e, f). The decrease of $S(q_{\rm max})$ at large $\phi$ is a distinctive feature of
ultrasoft potentials~\cite{graves2005,ZhangNat2009}.
It is ascribed to the
entropy gained by the exploration of a large number of disordered
configurations whose energy cost remains modest due to the soft
particle interaction~\cite{jacquin}.
This anomalous structural evolution
suggests that the dynamical hallmarks in regime III are specific to soft colloids, in contrast to those of regime II, which are not.

\begin{figure}
 \includegraphics[width=1\columnwidth,clip]{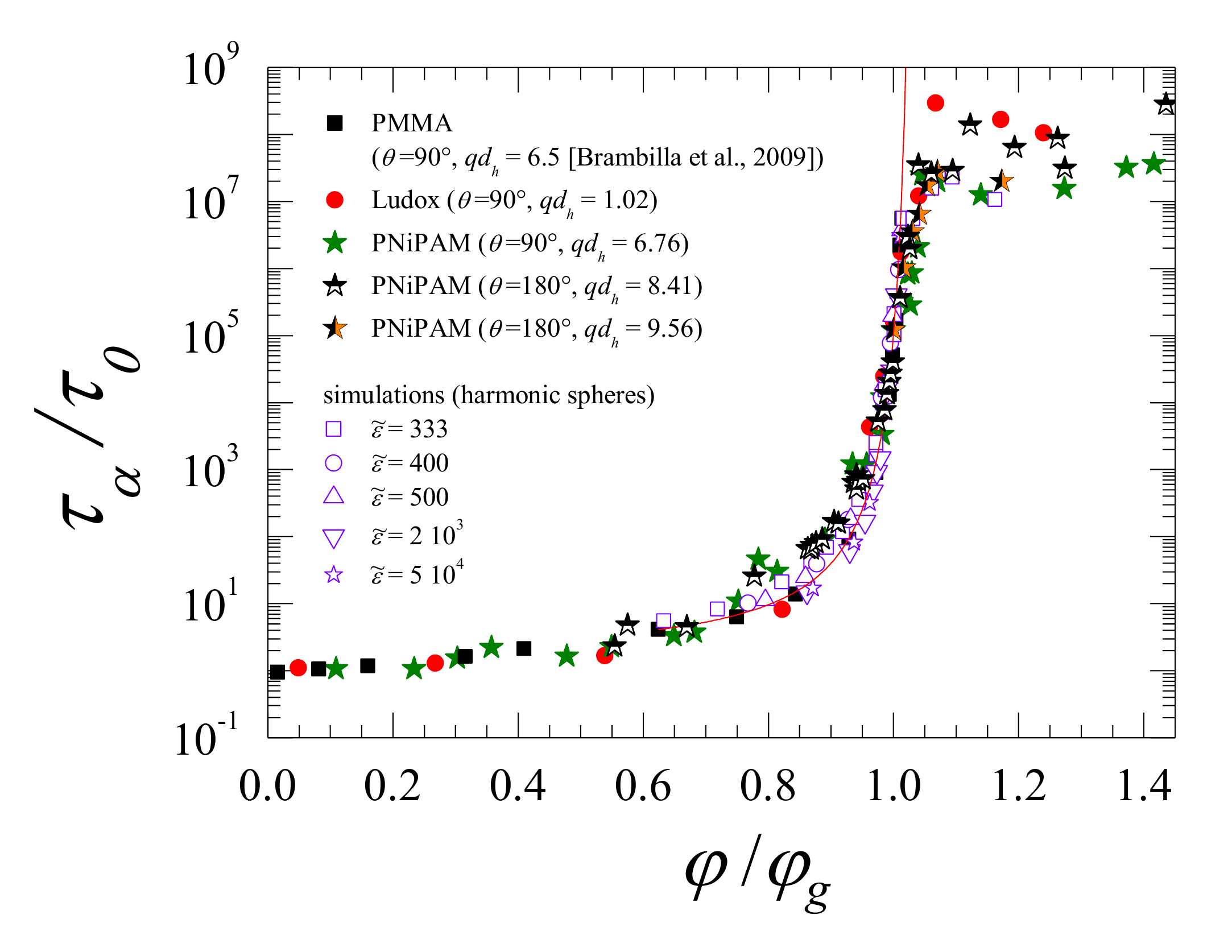}
 \caption{Rescaled relaxation time versus volume fraction
rescaled by the operational glass transition $\phi_g$.
Solid symbols are experiments on soft particles (PNiPAM and Ludox)
and PMMA hard spheres (data taken from \cite{BEPPSBC08}).
Open symbols: simulations of harmonic spheres whose adimensional softness
$\tilde{\epsilon}$ is varied from the hard to the ultrasoft limit.
The line is a fit to data in regime II according to $\tau_\alpha \sim
\exp[A/(\phi_0-\phi)]$, with $\phi_0 = (1.04 \pm 0.05) \phi_g$. 
}
 \label{fig:tau_vs_phi}
\end{figure}

This expectation is quantitatively confirmed in
Fig.~\ref{fig:tau_vs_phi} which gathers
$\tau_{\alpha}(\phi)$ for several  systems.
To compare different materials,
we follow the glass literature~\cite{BB11}
and rescale $\phi$
by $\phi_g$, a `glass transition' volume fraction arbitrarily defined by $\tau_\alpha(\phi_g)/\tau_0 = 10^5$,
which corresponds to $\tau_\alpha \approx 100$s for our systems.
We also display numerical data for soft harmonic particles
with softness varying over more than two orders of magnitude, and literature data from experiments on colloidal hard spheres~\cite{BEPPSBC08}.
We find that the sharp increase of the relaxation time in \ludo{the equilibrium}
regime II is {\it unaffected by particle softness, by the interaction type,
by internal degrees of freedom, or by particle deformability.}
All data collapse onto a master curve,
which is well described by the same steep exponential divergence
describing the hard sphere behavior~\cite{BEPPSBC08}.
\ludo{Other \ludo{fragile} functional forms have been tested, yielding similar results~\cite{SupInfo}.}
 This behavior is also robust with respect to the probed lengthscale, since data collected at
various $qd_h$ fall onto the same mastercurve.
This universal behavior is in stark contrast with the central finding
of Ref.~\cite{mattsson2009}. Our PNiPAM microgels are slightly
softer~\cite{SupInfo} than the softest particles of~\cite{mattsson2009}; \ludo{thus, the discrepancy can not be attributed to particle softness itself}. \textcolor{myc} {Rather, we attribute it to osmotic deswelling, which is specific to charged microgels such as those of~\cite{mattsson2009}. Recent work ~\cite{pelaez-fernandez_impact_2015,van_der_scheer_fragility_2017} indicates that charged microgels significantly deswell as their concentration is increased, due to the decrease in the osmotic contribution of the counterions to particle swelling. Owing to deswelling, both the particle size and the interparticle interactions change with $\varphi$, resulting in the observed `strong' behaviour.
\ludo{Altogether, the idea that softness alone affects the nature of the
growth of $\tau_\alpha$ in the equilibrated supercooled regime needs drastic revision.}}

\begin{figure}
 \includegraphics[width=1\columnwidth,clip]{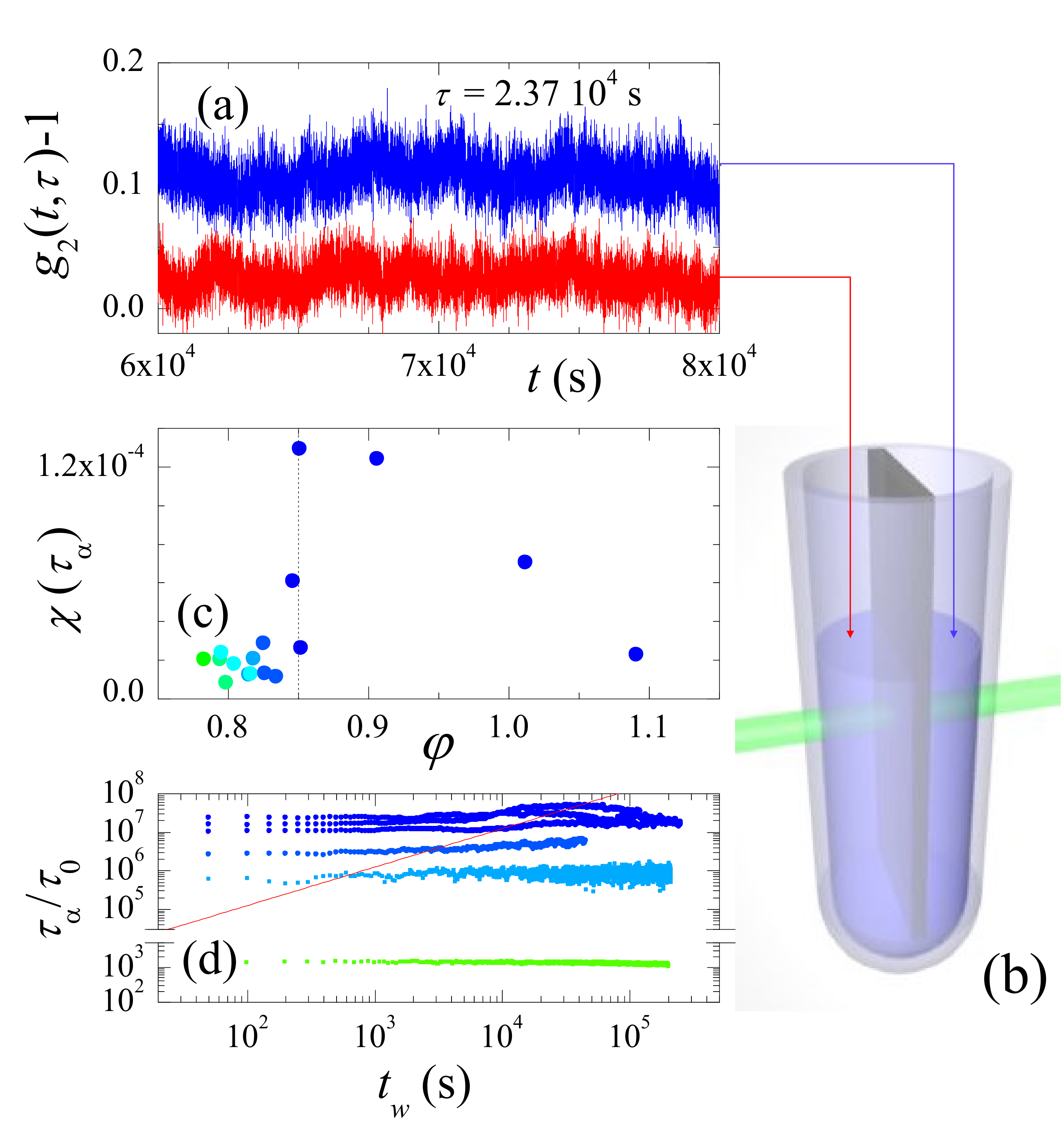}
 \caption{(a) Temporal fluctuations of the dynamics at a fixed
time lag $\tau \approx \tau_\alpha$, for a PNiPAM sample with $\phi = 0.851$
(data are offset vertically for clarity). The two curves are measured
simultaneously in two independent sample chambers in the setup
sketched in panel (b), and display uncorrelated fluctuations.
(c) Maximum of the dynamical susceptibility $\chi(\tau_\alpha)$
as a function of $\phi$ for the PNiPAM samples.
(d) Age dependence of the dynamics for representative
PNiPAM samples. From bottom to
top, $\phi = 0.743$, 0.818, 0.835, 1.012, 1.091, 0.906. \textcolor{myc} {The line shows $\tau_\alpha = t_w$: in all cases, we access the $\tau_\alpha < t_w$ regime where the structural relaxation and aging time scales are well separated.}}
\label{fig:2comp}
\end{figure}

While regime II is generic to all colloids \ludo{(hard and soft)},
the behaviour in regime III is \ludo{instead} specific to very soft colloids; it is not observed for hard spheres. An explanation for the mild $\phi$ dependence of $\tau_\alpha$ in regime III could be that measurements are limited by setup instabilities. To rule this out we performed several tests~\cite{SupInfo}, including the simultaneous measurement of the dynamics of the same system placed in two independent chambers, as shown in Fig.~\ref{fig:2comp}(b)~\cite{duri_resolving_2009,cipelletti_simultaneous_2013}. The fluctuations in $g_2 - 1$ observed for the two chambers are not correlated, see Fig.~\ref{fig:2comp}(a). \ludo{Therefore, the observed dynamical fluctuations do not result from instabilities of the experimental setup, but occur spontaneously within the sample itself.}

\ludo{The dynamical behaviour in regime III is clearly not the smooth continuation of the equilibrium regime II.
Our experiments suggest that the system is in fact out of equilibrium and displays aging behaviour. In Fig.~\ref{fig:2comp}(d), we show the age dependence of $\tau_\alpha$ for representative concentrated PNiPAM samples. At $\phi =0.743$, in regime II, $\tau_\alpha$ is age-independent, as expected for equilibrium dynamics. At higher $\phi$, in regime III, the relaxation time fluctuates erratically and increases, albeit very slowly, suggesting lack of full equilibration. Note that for all samples we access the $t_w > \tau_\alpha$ regime (see red line in Fig.~\ref{fig:2comp}(d)), which insures that the experiments lasted long enough for the measured relaxation time not to be limited by the aging time itself.

The interpretation of regime III as an aging regime is further confirmed by the appearance of a compressed exponential decay of time correlation functions ($\beta \ge 1$), and by the significant temporal fluctuations of the dynamics observed experimentally in regime III (see Fig.~\ref{fig:2comp}(a)). These are typical signatures found in non-equilibrium, glassy soft matter, which result from an intermittent release of internal stresses~\cite{Cipelletti2003,madsen_beyond_2010}. These events are known to trigger long-ranged dynamical heterogeneities~\cite{duri_resolving_2009,maccarrone_ultra-long_2010,lieleg_slow_2011,agnes_epl}. We have indeed detected a sharp increase of dynamical heterogeneity at the transition between regimes II and III. This is shown in Fig.~\ref{fig:2comp}(c), which reports the evolution of the dynamical susceptibility $\chi(\tau_\alpha) = \mathrm{var}[g_2(t,\tau_\alpha)]$~\cite{Duri2005,BBBCS2011}. An analogous non-monotonic behavior of the dynamic susceptibility was reported previously for similar concentrated PNiPAM suspensions~\cite{rahmani_dynamic_2012}.

A key difference between the aging of hard and soft particles is the very mild increase of the relaxation time with both sample age and packing fraction in Figs.~\ref{fig:tau_vs_phi} and \ref{fig:2comp}(d). By contrast, spontaneous relaxation cannot be observed in hard sphere glasses. We believe that the difference stems from the observed structural evolution of soft systems, which display a weakening of local order, as revealed by the sharp decrease of $S(q_{\rm max})$ at large $\phi$. In glassy systems, increasing the structural disorder typically accelerates the dynamics~\cite{berthier2010pre}, which indeed appears to be faster in soft spheres than it is in hard spheres, for which $S(q_{\rm max})$ keeps increasing with $\phi$ in the glass phase.}

Our work shows that the glassy dynamics of soft colloids
is markedly different from what has been assumed so far. Regardless
of their softness, colloids
exhibit in the supercooled regime a sharp increase of the
equilibrium relaxation time similar to that of hard colloids.
But in contrast to hard spheres, at larger density
soft colloids enter a peculiar aging regime characterized
by intermittent release of internal stress and \ludo{with a mild aging of the structural relaxation time, due to particle softness.}
\textcolor{myc} {Our work provides a coherent picture
of the glass transition of colloidal particles interacting via a soft potential. The strong similarities between the dynamics of charged compact particles and microgels suggests that the behavior uncovered by our experiments is quite general and insensitive to the details of the interaction potential. The comparison with previous findings for polymer-based systems and charged microgels~\cite{vlassopoulos_multiarm_2001,pelaez-fernandez_impact_2015,van_der_scheer_fragility_2017} indicates that \ludo{it is not softness \textit{per se }but other mechanisms, such as osmotic deswelling and polymeric degrees of freedom, which are likely responsible for the `strong'-like behavior reported earlier for some of these systems.}}

\begin{acknowledgments}
We thank A. Fernandez-Nieves and J. Mattsson for useful discussions. The research leading to these results has
received funding from CNES, the Swiss National Science Foundation (200020\_130056), the European Research Council
under the European Union’s Seventh Framework
Programme (FP7/2007-2013)/ERC Grant Agreement No. 306845, and was supported by a grant from the Simons Foundation
(\# 454933, Ludovic Berthier)

\end{acknowledgments}


%

\newpage
\textbf{Supplemental Material}

\section{Particle size}

Two distinct batches of Poly-N-isopropylacrylamide (PNiPAM) microgels were synthesized by emulsion polymerization according to the protocol described in \cite{SenffJCP1999} and were suspended in a $2$ mM $\mathrm{NaN_3}$ aqueous solution, to prevent bacterial growth. PNiPAM microgels exhibit a lower critical solution temperature (LCST) close to room temperature, which results in a temperature-dependent size of the particles. We characterized the $T$-dependent size of our microgels by measuring their hydrodynamic diameter $d_h$ with conventional dynamic light scattering (DLS) \cite{BernePecora1976}, using a goniometer equipped with a Brookhaven correlator (BI-9000 AT) and working at very high dilution (w/w concentration $c  \le 10^{-5}$).
Most experiments reported in the main text were performed with batch 1, for which the hydrodynamic diameter ranges from $d_h = 282.8~\mathrm{nm}$ at $T = 303.15~\mathrm{K}$ to $d_h = 362.2~\mathrm{nm}$ at $T = 290.15~\mathrm{K}$. The data at $qd_h = 8.41$ shown in Fig. 3 of the main text were collected using batch 2, for which $d_h = 212~\mathrm{nm}$ at $T = 303.15~\mathrm{K}$ and $d_h = 268~\mathrm{nm}$ at $T = 293.15~\mathrm{K}$.

The silica particles are Ludox TM-50 by Sigma-Aldrich, suspended in salt-free Milli-Q water. Their size was measured by DLS in the very dilute limit ($c=10^{-6}$ w/w), finding $d_h=46$ nm.

\section{Sample volume fraction}
The quoted volume fraction of the microgel suspensions has been obtained from their mass fraction $c$ (known from the synthesis) using $\varphi = \alpha c$ , where $\alpha = 28.65$ at $T = 293$ K is determined by fitting the zero-shear viscosity of diluted suspensions to Einstein's law, $\eta=\eta_s(1+2.5\alpha c)$, with $\eta_s$ the viscosity of the solvent~\cite{Truzzolillo2015}. In the experiments, the volume fraction is varied either by changing $c$ or by varying $T$ and hence the particle size~\cite{SenffJCP1999}. Temperature is always kept in the range $290.15~\mathrm{K} \le T \le 303.15~\mathrm{K}$, well below the LCST, thus avoiding the onset of attractive interactions and the sharp change of electrophoretic mobility reported for PNiPAM microgels at high $T$~\cite{Daly2000}.

For the silica particles, $\varphi$ is controlled by varying $c$ and is calculated from $\varphi = c\rho_s/\left[ \rho_p-c(\rho_p-\rho_s)\right]$, where $\rho_s=0.998$ g/ml is the density of water at $20$ $^{\circ}C$  and $\rho_p=2.2$ g/ml that of the particles. $c$ is measured by drying an aliquot of the sample.

\section{Debye length of the Ludox suspensions}\label{Debye}

The screening length characterizing Ludox suspensions in absence of added salt is a function of the particle concentration and can be calculated by imposing that each ionized group on the particle surface releases one Na$^+$ counterion. We calculate the number of ionized groups per particles by measuring their $\zeta$-potential at very low concentration ($\varphi=10^{-4}$). We obtain $\zeta=-37 \pm 4$ mV. The measured $\zeta$-potential is the value of the electrostatic potential at the slipping plane, which is at a distance $\delta$ from the particle surface. $\zeta$ can be related to the surface potential $\Psi(R_h)$, i.e. the electrostatic potential at distance $R_h$ from the center of the colloid, via the Gouy-Chapman model \cite{Hunter1988,Sennato2012} as

\begin{equation}\label{GC-model}
    \Psi(R_h)=\frac{4k_BT}{e}\tanh^{-1}\left[\exp\left(\frac{\delta}{\xi_D}\right)\tanh\left(\frac{\zeta e}{4 k_B T}\right)\right]
\end{equation}
where $k_B$ is the Boltzmann constant, $T$ is the absolute temperature, $e$ the elementary charge, $\epsilon=6.90\cdot 10^{-10}$ F/m is the dielectric permittivity of water, and $\xi_D=[e^2\varphi z_p/(4/3\pi R_h^3\epsilon k_b T )]^{-1/2}$ is the Debye length that takes into account the counterions released by each particle in a suspension at particle volume fraction $\varphi$. By imposing that the electrostatic potential at the particle surface is $\Psi(R_h)=\frac{z_p e}{4\pi\epsilon R_h}$ and knowing that $0$ nm $\leq \delta \leq 0.25$ nm \cite{Kobayashi2014}
we calculate numerically the valence $z_p$ of each particle and hence the Debye length as a function of $\varphi$. In the range of volume fraction investigated here, we obtain $17$ nm$ \leq \xi_D \leq$ $24$ nm. It's worth noting that for volume fractions ranging from $\varphi=0.35$ (lower bound of Regime II) up to $\varphi=0.44$ (maximum volume fraction investigated in regime III) the Debye length varies only marginally, from $\xi_D=17$ nm up to $\xi_D=19$ nm. Therefore, the interaction potential is almost independent of $\varphi$ in the range of volume fractions over which the dynamics change the most.

\section{PniPam microgel softness}
We follow the method of~\cite{sierra-martin_determination_2011} and obtain the compression modulus of our PNiPAM particles by measuring the variation of the particle size upon applying an external osmotic pressure, which is imposed by adding polyethylene glycol (PEG) with molecular weight $M_w=35$ kDa. We measure the PNiPAM size using dynamic light scattering for increasing concentrations $c_{PEG}$ of polymer. To extract the microgel size from the DLS data, we use the viscosity $\eta_{\mathrm{PEG}}$ of the PEG35k solutions as measured by standard rheometry (for $c_{\mathrm{PEG}} \geq 1.5~\mathrm{wt}~\%$) or using an Anton Paar Lovis 2000 ME microviscosimeter (at lower $c_{\mathrm{PEG}}$). The osmotic pressure $\Pi$ has been measured with a cryo-osmometer (Gonotec - Osmomat 3000) for $\Pi>3000$ Pa. For lower pressures, i.e. low polymer concentrations, we obtain $\Pi$ from static light scattering (SLS) measurements. The isothermal compressibility of a PEG solution, $1/\chi_T=-c_{\mathrm{PEG}}(\frac{d\Pi}{dc_{\mathrm{PEG}}})$, is related to the low-$q$ limit of the intensity $I$ scattered by the polymer solution by~\cite{Danner1993}
\begin{equation}\label{compress}
1/\chi_T= \Gamma\frac{1}{I(q\Delta r \rightarrow 0)}\frac{4\pi^2c_{\mathrm{PEG}}^2n^2k_BT}{\lambda^4}\left(\frac{dn}{dc_{\mathrm{PEG}}}\right)^2,
\end{equation}
where $\lambda$ is the laser wavelength, $n$ the solution refractive index, $\Gamma$ a setup-dependent constant, and $\Delta r$ the distance between two monomer units. Note that in the range of $q$-vectors accessible by our setup we easily probe the $q\Delta r  \ll 1$ regime.
The dependence of $\frac{d\Pi}{dc}$ on polymer concentration was fitted by a quadratic polynomial. The fitting function was then integrated, yielding the $c_{\mathrm{PEG}}$-dependent osmotic pressure $\Pi$, up to the multiplicative constant $\Gamma$. $\Gamma$ was determined by matching the osmotic pressure determined from SLS data to that obtained by cryo-osmometry, in the range of $c_{\mathrm{PEG}}$ where both techniques are available.

\begin{figure}[htbp]
\includegraphics[width=1.\columnwidth]{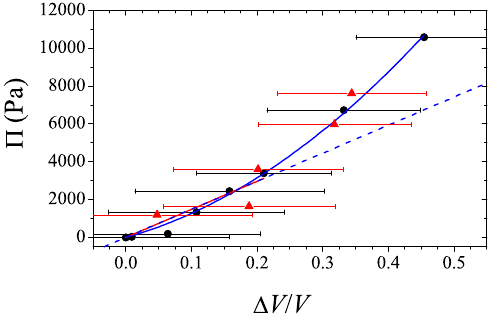}\\
Fig. SM-1: Osmotic pressure as a function of the relative variation of the microgel volume, $\Delta V/V$. Black points refer to the microgels used in this work, red triangles are data for the softest microgels of Ref.~\cite{mattsson2009}.
\label{compress}
\end{figure}

Figure SM-1 shows that, within the experimental uncertainty, our data closely match those reported by Mattsson et al.~\cite{mattsson2009} for their softest microgels. We calculate the compression modulus $K=-V(\frac{d\Pi}{dV})$ of our PNiPAM microgels by performing a linear fit of the data at low pressure ($\Pi<4000$ Pa, dashed line in Fig. SM-1), obtaining $K=14.8$ kPa. Alternatively, we fit the data over the full range of available $\Delta V/V$ using a parabolic function (solid line), $\Pi=K\frac{\Delta V}{V}+K_2\left(\frac{\Delta V}{V}\right)^2$, obtaining $K=9.7$ kPa. We conclude that the compression modulus of our microgels is of the order of 10-15 kPa, indicating that our PNiPAM microgels are somehow softer than the softest particles of Ref.~\cite{mattsson2009}, for which $K=20$ kPa.

\section{Stability of the DLS setup}
We successfully established the apparatus stability by using a fully frozen sample. We use a frosted glass, for which unavoidable mechanical instabilities eventually lead to a spurious relaxation of $g_2-1$, but only on very long time scales $\ge 3.5 \times 10^6~\mathrm{s}$, more than a factor of 3.75 (resp., $> 10$ times) longer than the longest relaxation time measured for the silica (resp. the PNiPAM) particles. In addition, by collecting the scattered light in an imaging geometry~\cite{duri_resolving_2009,cipelletti_simultaneous_2013}, we are able to detect a potential drift of the sample that could also spuriously accelerate the dynamics, as recently argued by Gabriel and coworkers~\cite{gabriel_compressed_2015}. We find no measurable drift, confirming the setup stability.

\section{Rheology}
The flow curves for all samples in the regimes II and III were obtained by performing steady rate rheology experiments, using a stress-controlled AR 2000 rheometer (TA Instruments), equipped with a steel cone-and-plate geometry (cone diameter = 25 mm, cone angle = 0.1 rad). For the samples in the regime I a bigger cone (cone diameter = 50 mm, cone angle = 0.0198 rad) has been used.

\section{Static structure factor $S(q)$}
Figure SM-2 shows representative static structure factors $S(q)$ measured for PNiPAM (a) and Ludox (b) suspensions, at various $\varphi$. Note that the height $S(q_{max})$ of the first maximum of $S(q)$ initially increases with $\varphi$, but eventually \textit{decreases} upon increasing volume fraction, as discussed in the main text. The height of the peak, $S(q_{max})$, reported in Fig. 2e-f) of the main text is obtained from a Lorentzian fit (lines in Fig. SM-2).
\begin{figure}[htbp]
\includegraphics[width=1.\columnwidth]{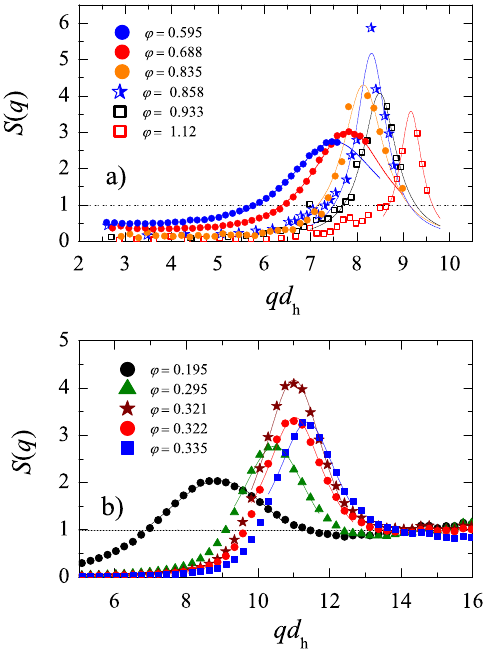}\\
Fig. SM-2: Representative static structure factors $S(q)$ for PNiPAM (a) and Ludox (b) suspensions. Lines are Lorentzian fits to the peak of $S(q)$.
\end{figure}

\section{Additional data for Ludox suspensions}

\subsection{Rheology and dynamic light scattering}
Analogously to what has been reported in Figure 1 of the main text for the PNiPAM microgels, we show below in Fig.~SM-3 the flow curves and the intensity correlation functions of suspensions of Ludox particles at representative volume fraction $\varphi$.

\begin{figure}[htbp]
\includegraphics[width=1.\columnwidth]{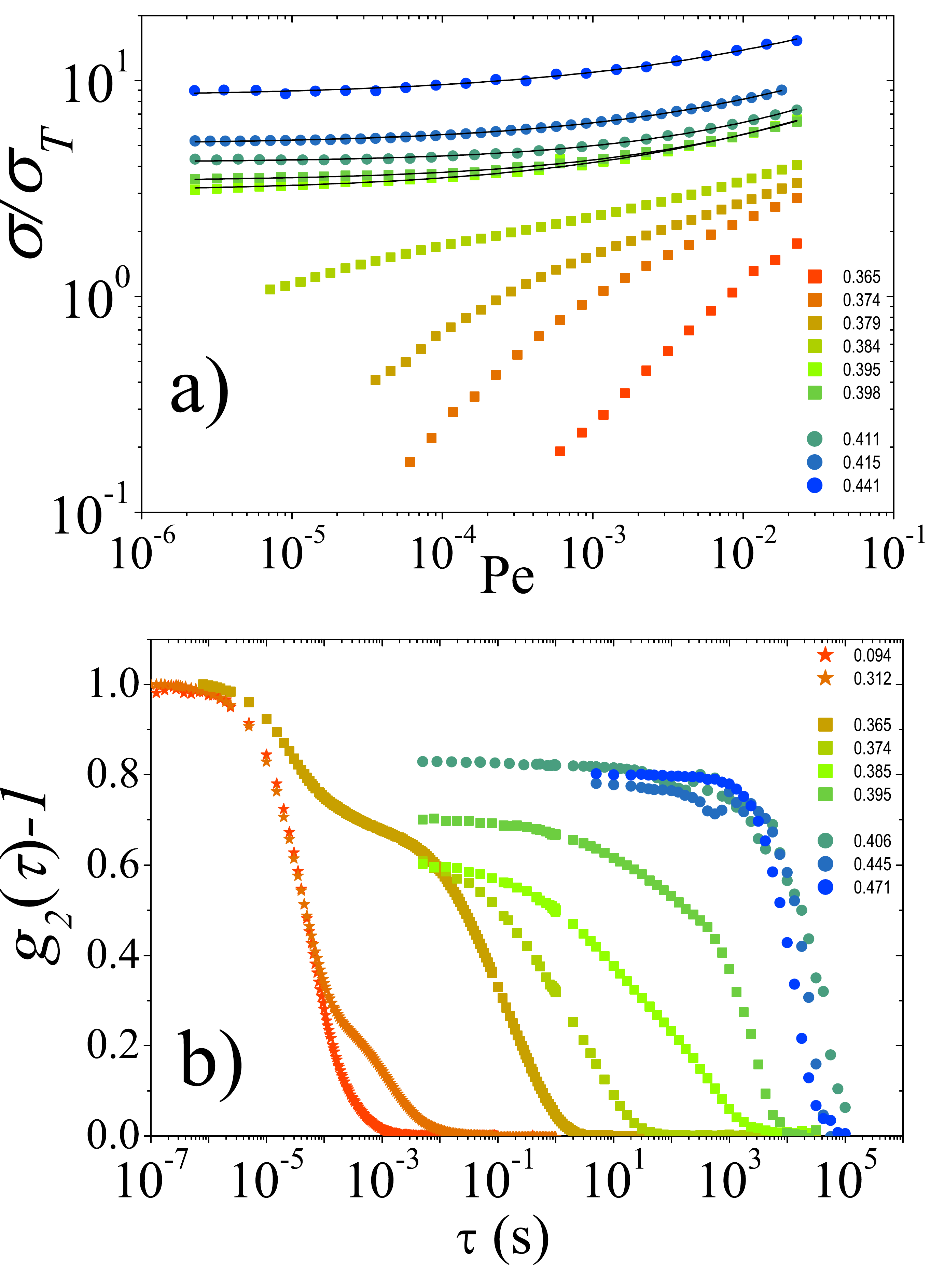}\\
Fig. SM-3: Selected flow curves for Ludox suspensions. Curves are labelled by the volume fraction; the lines are
Herschel-Bulkley fits. b): Representative correlation functions for Ludox suspensions. In both panels, star, square and circle symbols correspond to concentration regimes I, II and III, respectively, as in Fig. 1 of the main text.
\label{flow-g2}
\end{figure}

\subsection{Aging}
Figure SM-4 shows the age dependence of $\tau_{\alpha}$ for representative concentrated Ludox samples. Note the fluctuations of $\tau_{\alpha}$, which indicate that the suspensions are not fully equilibrated.

\begin{figure}[htbp]
\includegraphics[width=1.\columnwidth]{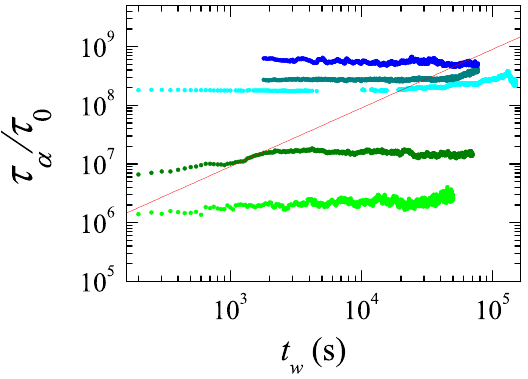}\\
\label{ageing-Ludox} Fig. SM-4: Age dependence of the dynamics for representative Ludox suspensions. From top to bottom: $\varphi=0.471$, $\varphi=0.445$, $\varphi=0.406$, $\varphi=0.395$, $\varphi=0.385$. \textcolor{myc} {The line shows the $\tau_\alpha = t_w$ function: for all samples, the experiment duration is long enough to access the regime $\tau_\alpha < t_w$, where the relaxation time is not dictated by the sample age}.
\end{figure}

\begin{figure}[htbp]
\includegraphics[width=1.\columnwidth]{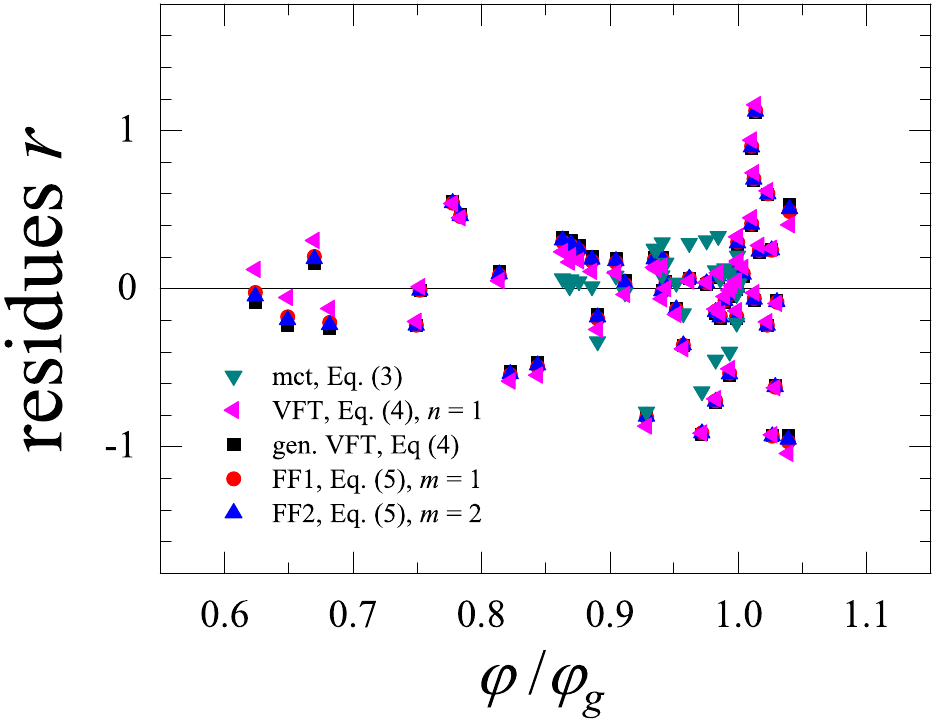}\\
\label{fig:fits} \textcolor{myc} {Fig. SM-5: Residues of fits with various functional forms to the $\log \left (\tau_\alpha/\tau_0 \right )$ vs $\varphi/\varphi_g$ data of Fig. 3 of the main manuscript. Only data in the supercooled regime, Regime II, have been fitted. See text for more details.}
\end{figure}

\textcolor{myc} {
\section{Fits to the data in the supercooled regime, Regime II}}
\textcolor{myc} {
We have tested various fitting functions to model the data of Fig. 3 of the main manuscript in Regime II (supercooled regime). We fit $\log \left (\tau_\alpha/\tau_0 \right )$ as a function of $\phi = \varphi/\varphi_g$ with the following expressions:
\begin{equation}\label{mct}
    \log \left (\tau_\alpha/\tau_0 \right ) = -\gamma \log(\phi_{mct}-\phi) + B   \,,
\end{equation}
\begin{equation}\label{VFT}
    \log \left (\tau_\alpha/\tau_0 \right ) = \frac{A}{(\phi_{0}-\phi)^n} + B   \,,
\end{equation}
\begin{equation}\label{FF1and2}
    \log \left (\tau_\alpha/\tau_0 \right ) = \frac{A}{(\phi_{0}-\phi)^{m-1}}\exp\left [ \frac{C}{(\phi_{0}-\phi)^{m}} \right ] + B  \,\,\,\,\, m = 1, 2\,.
\end{equation}
Equation~\ref{mct} is the mode coupling theory (mct) prediction. Equation~\ref{VFT} with $n=1$ is the usual Vogel-Fulcher-Tammann law~\cite{debenedetti_supercooled_2001}. Leaving $n$ as a free fitting parameter yields a generalized VFT law, analogous to the Avramov equation proposed for molecular glass formers~\cite{hecksher_little_2008-1} with the substitution $1/T \rightarrow 1/(\phi_{0}-\phi)$, to account for the apparent divergence of the relaxation time as the colloidal volume fraction approaches a critical packing fraction $\phi_{0}$. Equation~\ref{FF1and2} is the colloidal counterpart of the functions FF1 and FF2 (for $m=1$ and 2, respectively) proposed in~\cite{hecksher_little_2008-1}, where again we have substituted $1/T$ with $1/(\phi_{0}-\phi)$. Figure SM-5 shows the residues $r = \log \left (\tau_\alpha/\tau_0 \right )_{data}-\log \left (\tau_\alpha/\tau_0 \right )_{fit}$ as a function of $\varphi/\varphi_g$. The experimental and simulation relaxation times have been fitted in the range $0.6 \le \log \tau_\alpha/\tau_0 \le 7.0$, except for mct, where a smaller fitting range had to be used, $2.1 \le \log \tau_\alpha/\tau_0 \le 5.1$. Attempts to extend the mct fitting range led to nonphysical values of the exponent $\gamma$, which mct predicts to be in the range $2.5-2.7$
: as reported previously~\cite{brambilla_comment_2010-1}, the mct exponent rapidly increases beyond $\gamma= 2.7$ when data at higher volume fractions are included in the fit. As seen in Fig. SM-5, all fits give very close results. We quantify the goodness of the fits by calculating $\tilde{\chi}^2$, the reduced chi-squared defined as $\tilde{\chi}^2 = (n_p-p)^{-1} \sum_{i=1}^{n_p} [\log \left (\tau_{\alpha,i}/\tau_0 \right )_{data}-\log \left (\tau_{\alpha,i}/\tau_0 \right )_{fit}]^2$, with $n_p$ the number of fitted data points and $p$ the number of fitting parameters. The reduced chi-squared is shown in Table SM-T1, together with the values of the fitting parameters. All functions have a very similar reduced chi-squared, indicating equal quality fits, with the exception of the mct expression, which has a smaller $\tilde{\chi}^2$. However, this is a consequence of the reduced interval over which the mct fit has been performed: when fitting the same set of data points with the other expressions, one recovers essentially the same fit quality. For example, over the reduced range of relaxation times used for the mct fit, one finds for the VFT expression $\tilde{\chi}^2 = 0.079$, to be compared to 0.077 for the mct function. In the main manuscript, we choose the VFT expression, since it fits satisfactorily the largest range of data with the smallest number of parameters.
}
\begin{table}
\textcolor{myc} {
TABLE SM-T1: fitting parameters and reduced chi-squared for various fitting functions, in the supercooled regime (see text for more details). The fitting range is $2.1 \le \log \tau_\alpha/\tau_0 \le 5.1$ for mct and $0.6 \le \log \tau_\alpha/\tau_0 \le 7.0$ for all the other functions.\\
\begin{ruledtabular}
\begin{tabular}{cccccccc}
Function &$\phi_0, \phi_{mct}$ & $A$ &$B$ & $\gamma$ & $n$ & $C$ & $\tilde{\chi}^2$\\
\hline
mct&1.01&&-0.19&2.71&&&0.077 \\
VFT&1.10&0.55&-0.48&&1 (fixed)&&0.127 \\
generalized VFT&1.10&0.61&-0.59&&0.94&&0.129\\
FF1&1.13&6.08&-6.52&&&0.08&0.129\\
FF2&1.19&0.69&-0.63&&&0.015&0.129\\
\end{tabular}
\end{ruledtabular}
}
\end{table}

\bibliographystyle{apsrev}


\end{document}